\documentclass[twocolumn,english,showpacs,preprintnumbers,amsmath,amssymb,floatfix]{revtex4}
\usepackage[T1]{fontenc}
\usepackage[latin9]{inputenc}
\usepackage{color}
\usepackage{array}
\usepackage{amstext}
\usepackage{graphicx}
\usepackage{epstopdf}
\usepackage{esint}

\usepackage{lmodern}
\makeatletter

\providecommand{\tabularnewline}{\\}

\@ifundefined{textcolor}{}
{%
 \definecolor{BLACK}{gray}{0}
 \definecolor{WHITE}{gray}{1}
 \definecolor{RED}{rgb}{1,0,0}
 \definecolor{GREEN}{rgb}{0,1,0}
 \definecolor{BLUE}{rgb}{0,0,1}
 \definecolor{CYAN}{cmyk}{1,0,0,0}
 \definecolor{MAGENTA}{cmyk}{0,1,0,0}
 \definecolor{YELLOW}{cmyk}{0,0,1,0}
 }

\@ifundefined{definecolor}
 {\usepackage{color}}{}
\@ifundefined{definecolor}
 {\usepackage{color}}{}
\makeatother

\makeatother

\usepackage{babel}

\begin{document}

\title{Target mass corrections and higher twist effects in polarized deep-inelastic scattering}

\author{S. Taheri Monfared $^{a    }$}
\email{Sara.taheri@ipm.ir}

\author{Z. Haddadi $^{b}$}
\email{z.haddadi@rug.nl}

\author{Ali N. Khorramian $^{c,a}$}
\email{Khorramiana@theory.ipm.ac.ir}

\affiliation{$^{(a)}$ School of Particles and Accelerators, Institute for
Research in Fundamental Sciences (IPM), P.O.Box 19395-5531, Tehran, Iran \\
 $^{(b)}$ KVI/University of Groningen, 9747 AA Groningen, Netherlands\\
 $^{(c)}$ Physics Department, Semnan University, Semnan, Iran}

\date{\today}
\begin{abstract}
We perform a next-to-leading-order QCD analysis to world data on polarized structure functions $g_1$ and $g_2$ in a fixed-flavor number scheme.
We include target mass corrections and higher twist effects in our fitting procedure and study their non-negligible effects on physically interesting quantities.
Twist-3 contributions to both polarized structure functions are determined, and the accuracy of the extracted polarized parton distribution functions is improved.
$^{3}\rm He$ and $^{3}\rm H$ polarized structure functions are described based on our fit result. Moreover, sum rules are derived and compared with available theoretical and experimental results.
\end{abstract}


\maketitle
\tableofcontents{}
\section{Introduction}
The determination of the nucleon's spin into its quark and gluon components is still an important challenge in particle physics.
The deep-inelastic scattering (DIS) experiments performed at DESY, SLAC, CERN, and JLAB have refined our understanding of the spin distributions and revealed the spin-dependent structure functions of the nucleon. The polarized structure functions $g_1(x,Q^2)$ and $g_2(x,Q^2)$ are measured in deep-inelastic scattering of a longitudinally polarized lepton on polarized nuclear targets.

Theoretical models have remarkably improved since the early framework of quark parton model (QPM) indicated that $g_1$ measures only quark contributions to the  nucleon's spin and $g_2$ is identically zero. Afterward, perturbative quantum chromodynamics (pQCD) analysis in next-to-leading-order (NLO) approximation provides information on the role of gluons in the overall spin of the nucleon.
Moreover, $g_2$ contains nonperturative higher twist (HT) contributions, such as quark-quark and quark-gluon correlations and quark mass effects, which are not interpreted in QPM. Operator Product Expansion (OPE) based on QCD is an appropriate formalism that is applicable to interpret $g_2$ structure function \cite{Abe:1997qk,JLABn2005}.
\par
The traditional method to perform global fits concentrates on the extraction of leading twist parton distribution functions (PDFs), using cuts on minimum values of $Q^2$ and hadronic final-state mass squared $W^2$. The cuts are of the order $Q^2 \gtrsim 4$ GeV$^2$ and $W^2 \gtrsim 14$ GeV$^2$ \cite{oai:arXiv.org:0901.0002,oai:arXiv.org:1302.6246}, which means that $x$ is limited to $x \lesssim 0.7$. These kinematic cuts eliminate the contribution of corrections from various nonperturbative effects at finite $Q^2$, such as target mass corrections (TMCs) \cite{Schienbein:2007gr} and dynamical higher twist contributions \cite{Simula:2001iy}. These corrections  become increasingly significant as $Q^2$ is reduced and $x$ tends to $1$. The large-$x$ behavior of PDFs is extrapolated in this classical scenario.

Extraction of polarized PDFs (PPDFs) from a variety of experiments within NLO analyses is an important phenomenological issue \cite{Ball:2013hta,Jimenez-Delgado:2013boa,Leader:2010rb,deFlorian:2008mr,Blumlein:2010rn,Hirai:2008aj,Bourrely:2001du,Khorramian:2010qa,Arbabifar:2013tma}.
To compensate for the scarcity of polarized high-energy data points available to global PPDF analyses, the applied cuts were relaxed; thereby, one is typically forced to make use of the data at lower $Q^2$ and higher $x$. Consequently,
pQCD calculation cannot be trusted alone. As will be shown subsequently, in this kinematical region ($Q^2 \sim 1-5~\rm  GeV^2,~4~\rm GeV^2 < W^2 < 10~\rm GeV^2$), TMCs and HT contributions are important.
Some of the existing studies, such as Refs. \cite{Jimenez-Delgado:2013boa,Leader:2010rb,Blumlein:2010rn,Leader:2009tr,Leader:2006xc,Leader:2002ni}, use these effects in their global fitting procedure. Note that the $g_2$ polarized structure function is not considered in these analyses. But it helps to determine the low-$Q^2$ corrections due to following reasons. First, leading and higher twist contributions appear with the same order of importance. Second, $g_2$ data are mostly in the low-$Q^2$ region in which the effects of TMCs and HT become significant.
Because of the mentioned points, we believe that, although data for spin structure function $g_2$ is not accurate enough, at the current level of accuracy, analyzing both $g_1$ and $g_2$ structure functions provides a fertile ground to study the mentioned effects.
A future high-luminosity machine, like the EIC, is required to study the twist-3 contributions in detail \cite{oai:arXiv.org:1108.1713}.

In our latest analysis \cite{Khorramian:2010qa}, we determined PPDFs based on Jacobi polynomials using only $g_1$ experimental data and simply considered $g_1(x, Q^2)=g_1(x, Q^2)_{\rm pQCD}$.
In the present study, we improve our precision with full $g_1$ and $g_2$ analysis including TMCs and HT contributions. No polynomial technique is adopted.

The outline of the paper is as follows. We give an introduction to the theoretical framework that describes polarized structure functions in Sec. \ref{Theoretical analysis}.  Section \ref{QCD analysis and fitting procedure} provides detailed  information about our QCD analysis. A discussion of fit results is given in Sec. \ref{Discussion of fit results}. In Sec. \ref{Polarized Structure function of $^{3}He$ and $^{3}H$}, we compute the nuclear structure functions, and in Sec. \ref{Sum rules}, we check various polarized sum rules. Section \ref{Conclusions} contains the concluding remarks.

\section{Theoretical analysis}
\label{Theoretical analysis}
In the QCD polarized structure function, $g_1$ consists of two parts, the leading twist (LT)  ($\tau=2$) and the higher twist ($\tau \geq 3$) contributions:
\begin{equation}
g_1(x, Q^2) = g_1(x, Q^2)_{\rm LT} + g_1(x, Q^2)_{\rm HT}~.
\label{g1QCD}
\end{equation}
The LT term can be determined from
\begin{eqnarray}
g_1(x, Q^2)_{\rm LT}&=& g_1(x, Q^2)_{\rm pQCD}+ h^{\rm TMCs}(x,
Q^2)/Q^2 \nonumber\\
&+& {\cal O}(M^4/Q^4)~. \label{g1LT}
\end{eqnarray}
$g_1(x, Q^2)_{\rm pQCD}$ is achievable via NLO perturbative QCD when the nucleon mass is put equal to zero and  $h^{\rm TMCs}$  is calculable in pQCD. It is kinematic in origin and contains terms suppressed by powers of $M^2/Q^2$ at large values of $Q^2$.

The contribution of multiparton correlation in the nucleon is considered through the dynamical higher twist terms
\begin{equation}
g_1(x, Q^2)_{\rm HT}= h(x, Q^2)/Q^2 + {\cal O}(\Lambda^4/Q^4)~,
\label{HTQCD}
\end{equation}
where $h(x, Q^2)$ are nonpurtubative effects that can be calculated in a model-dependent manner.
They are dynamical in origin and suppressed by powers of  $\Lambda^2/Q^2$. $\Lambda$ is the scale of nonperturbative parton-parton correlation.
These corrections become increasingly important at low-energy scale.

The spin-structure function $g_2$ does not have a direct interpretation in pQCD. It can be understood using the OPE in which $g_2$ is separated into \cite{Anselmino:1994gn}
\begin{equation}
g_2(x,Q^2)=g_2^{WW}(x,Q^2)+\bar g_2(x,Q^2).
\label{WW}
\end{equation}
Here, $g_2^{WW}(x,Q^2)$ is a twist-2 part, and
\begin{equation}
\bar g_2(x,Q^2)=-\int _x ^1 \frac{\partial}{\partial y}\left[\frac{m_q}{M}h_T(y,Q^2)+\zeta(y,Q^2) \right]\frac{dy}{y}.
\end{equation}
The twist-3 part, $\zeta(y,Q^2)$, arises from nonperturbative multiparton interaction, which will be discussed in the next section.
 $~h_T$ depends on the quark transverse polarization density in twist-2, which is suppressed by the ratio of the quark to nucleon masses $\frac{m_q}{M}$.
Consequently, any deviation of $g_{2}$ from $g_{2}^{WW}$ is from the twist-3 contribution. 
It is in special properties of $g_2$ that its HT contribution can be equally important as its twist-2 part, since it is not suppressed by inverse powers of $Q^2$.
\subsection{Leading twist}
The leading twist contributions to the $g_1(x,Q^2)$ for the proton and neutron are available 
in the NLO \cite{Lampe:1998eu} by
\begin{eqnarray}
g_{1}(x,Q^2)_{pQCD} & = & \frac{1}{2}\sum\limits _{q}^{n_f}e_{q}^{2}\left\{[\delta q+\delta\bar{q}]\right.
  \otimes   (1+\frac{\alpha_{s}}{2\pi}\delta C_{q})\nonumber \\
 & + & \left.\frac{\alpha_{s}}{2\pi} \delta g \otimes \frac{\delta C_{g}}{n_f}\right\} \;.
 \end{eqnarray}
Here, typical convolution in $x$ space is represented with the symbol $\otimes$.
 $\delta q$, $\delta\bar{q}$, and $\delta g$ are polarized quark, antiquark, and gluon densities, which are evolved to $Q^2$ with the solution of DGLAP evolution equations in Mellin space. $\delta C_{q,g}$ are Wilson coefficient functions in NLO.
The deuteron structure function can be obtained via the relation
\begin{eqnarray}
g_{1}^d(x,Q^2)_{pQCD}&=&\frac{1}{2}\{g_1^{p}(x,Q^2)_{pQCD}+g_1^{n}(x,Q^2)_{pQCD}\}\nonumber \\
&\times&(1-1.5 w_D)~,
\end{eqnarray}
from proton and neutron ones, where $w_D=0.05\pm0.01$ is the probability to find the deuteron in a $D$state.
\par
Because of the fact that $g_1$ and $g_2$ contain the same twist-2 operators, the leading twist part of $g_2$ can be extracted via the Wandzura--Wilczek (WW) relation \cite{Wandzura:1977qf,Piccione:1997zh}
\begin{eqnarray}
g_{2}^{WW}(x,Q^{2})_{\rm pQCD}&=&-g_{1}^{p}(x,Q^{2})_{\rm pQCD}\nonumber\\
&+&\int_{x}^{1}\frac{dy}{y}g_{1}^{p}(y,Q^{2})_{\rm pQCD}~.
\label{eq:xg2}
\end{eqnarray}
 This relation remains valid when target mass corrections are included in the twist-2 contribution \cite{Piccione:1997zh,Accardi:2009au}.
\subsection{Target mass corrections and threshold problem}
To perform a reliable fit that contains data at lower values of $Q^2$, nucleon mass corrections cannot be neglected.
We follow the method suggested by Refs. \cite{Piccione:1997zh,Blumlein:1998nv,Nachtmann:1973mr}, which is exactly calculable and effectively belongs to the LT term \cite{Dong:2006jm}.

There is a traditional challenge with the behavior of the both polarized and unpolarized target mass corrected structure functions in the neighborhood of $x=1$.
Many attempts have been made to avoid this issue by considering various prescriptions in the literature \cite{Schienbein:2007gr,Accardi:2008pc,Georgi:1976ve,Piccione:1997zh,DeRujula:1976ih,Ellis:1982cd}.
These solutions are not unique.
In this paper, we follow the prescription of Ref. \cite{D'Alesio:2009kv} to avoid the threshold problem in the polarized structure function.
They impose the simplest probability for hadronization $\theta(x_{TH}-x)$. Here, the largest kinematically accepted amount of $x$ for inelastic scattering is $x_{TH}$, which is defined as
\begin{equation}
x_{TH}=\frac{Q^2}{Q^2+\mu(2M+\mu)}\ ,
\label{eq:theta}
\end{equation}
where $\mu$ should be the lowest mass particle that can  be produced in the process of interest.
We modified our polarized structure functions by multiplying them into the $\theta$ function.

\subsection{Higher twist}
Higher twist terms arising from long-range nonperturbative multiparton correlations contribute at low values of $Q^2$.
The BLMP model \cite{Braun:2011aw} made a step in developing a usable parametrization for phenomenological analysis.
It constructed HT distributions from convolution integrals of the light-cone wave functions by considering a simple model based on three valence quarks and one gluon with the total zero angular momentum.

Accordingly, we applied the parametrization form suggested by the BLMP model,
\begin{eqnarray}
g_{2}^{tw-3}(x)&=&A[ln(x)+(1-x)+\frac{1}{2}(1-x)^2]\nonumber\\
&+&(1-x)^3[B-C(1-x)+D(1-x)^2\nonumber\\
&-&E(1-x)^3]\ ,
\label{eq:g2HT}
\end{eqnarray}
in our initial scale and fit the coefficients to the data.
 We applied a nonsinglet evolution equation, since higher twist contributions are specially important in large-$x$ values. This approach is compared with exact evolution equations for the gluon-quark-antiquark correlation in Ref. \cite{Braun:2011aw}. They are practically equal.

By the integral relation of
\begin{eqnarray}
g_{1}^{tw-3}(x,Q^2)&=&\frac{4x^2M^2}{Q^2}[g_2^{tw-3}(x,Q^2)\nonumber\\
&-&2\int_x^1 \frac{dy}{y}g_2^{tw-3}(y,Q^2)]\ ,
\label{eq:g1HT}
\end{eqnarray}
the twist-3 part of different spin-dependent structure functions , $g_1^{tw-3}$ and $g_2^{tw-3}$, are related \cite{Blumlein:1998nv}.

\section{QCD analysis and fitting procedure}
\label{QCD analysis and fitting procedure}
\subsection{Parametrization}
We have adopted the following parametrization at the initial scale of $Q_{0}^{2}=4$ GeV$^{2}$ for $q=\{u_{v},d_{v},\bar{q},g\}$:
\begin{equation}
x\:\delta q(x,Q_{0}^{2})={\cal N}_{q}\eta_{q}x^{a_{q}}(1-x)^{b_{q}}(1+c_{q}x)\ .
\label{eq:parm}
\end{equation}
The normalization constants ${\cal N}_{q}$,
 \begin{equation}
{\cal N}_{q}^{-1}=\left(1+c_{q}\frac{a_{q}}{a_{q}+b_{q}+1}\right)\, B\left(a_{q},b_{q}+1\right)\ ,
\label{eq:norm}
\end{equation}
are selected such that $\eta_{q}$ are the first moments of the PPDFs.
$B(a,b)$ is the Euler beta function.
Considering SU(3) flavor symmetry, we have $\delta\overline{q}\equiv\delta\overline{u}=\delta\overline{d}=\delta s=\delta\overline{s}$.

The free unknown parameters provide a fit with a large degree of flexibility.
Some of our input parameters are subjected to constraints due to following reasons:
\begin{itemize}
\item
The first moments of the polarized valence quark densities can be related to $F$ and $D$ as measured
in neutron and hyperon $\beta$ decays \cite{PDG}. These constraints lead to the values of
$\eta_{u_{v}}=0.928\pm0.014$ and $\eta_{d_{v}}= -0.342\pm0.018$.
\item
 $c_{\bar{q}}$ and $c_{g}$ are set to zero due to the present accuracy of the data.
No improvement is observed in the fit with nonzero values of them.
\item
The $b_{\bar{q}}$ and $b_{g}$ parameters, which control the large-$x$ behavior of the polarized sea quarks and gluons, have large uncertainties in a region that is dominated by the valence distribution. We fixed them with the ratio of $b_{\bar{q}}/b_{g} \sim 1.6$, which is derived from the analogous unpolarized parameters.
\end{itemize}
The rest of parameters $\{A,B,C,D,E\}$ are the unknown higher twist parameters for to $g_{2,\{p,n,d\}}$ and consequently  $g_{1,\{p,n,d\}}$.
They are determined from a simultaneous fit to the all polarized structure function data of $g_1$ and $g_2$.

The parameters $\{\eta_{u_{v}},\eta_{d_{v}},c_{\bar{q}},c_{g}\}$ and the ratio of $b$ values are frozen in the first minimization procedure.
In the second minimization, we fix $\{b_{\bar{q}},b_{g},c_{u_{v}},c_{d_{v}}\}$  and $\{A,B,C,D,E\}$ as demonstrated in Tables~\ref{tab:result} and \ref{tab:g2twist3}. There are potentially nine unknown parameters in the fit, including $\alpha_{s}(Q_{0}^{2})$, which provide enough flexibility to have a reliable fit.
%
\begin{table}
\caption{\label{tab:result} Final parameter values and their statistical
errors at the input scale $Q_0^2=4$ GeV$^2$ determined from two different global analyses. Those marked with ($^*$) are fixed.}
\begin{tabular}{|c|c|c|c|}
\hline\hline
\multicolumn{2}{|c|}{Parameters}&{Full scenario}&{pQCD scenario}  \\
\hline\hline $\delta u_{v}$ & $\eta_{u_{v}} $ & $~0.928^*~$& $~0.928^*~$  \\
& $a_{u_{v}}$ & $0.558\pm 0.012$& $0.619\pm 0.018$  \\
& $b_{u_{v}}$ & $3.460\pm 0.006$& $3.234\pm 0.077$  \\
& $c_{u_{v}}$ & $8.848^*$& $5.468^*$ \\  \hline
$\delta d_{v}$ & $\eta_{d_{v}} $ & $-0.342^*$& $-0.342^*$  \\
& $a_{d_{v}}$ & $0.250\pm 0.033$& $0.226\pm 0.042$ \\
& $b_{d_{v}}$ & $3.912\pm 0.116$& $3.822\pm 0.357$ \\
& $c_{d_{v}}$ & $~14.162^*~$& $~25.09^*~$  \\ \hline
$~\delta_{\bar{q}}$ & $~\eta_{\bar{q}}$ & $-0.0605\pm 0.006$& $-0.0565\pm 0.022$  \\
& $a_{\bar{q}}$ & $0.567\pm 0.009$& $0.597\pm 0.075$   \\
& $b_{\bar{q}}$ & $4.993^*$ & $7.355^*$   \\
& $c_{\bar{q}}$ & $~0.0^*~$& $~0.0^*~$ \\ \hline
$\delta g$ & $\eta_{g} $ & $0.201\pm 0.044$& $0.147\pm 0.054$ \\
& $a_{g}$ & $2.253\pm 0.010$& $3.177\pm 0.58$  \\
& $b_{g}$ & $3.082^*$ & $4.540^*$ \\
& $c_{g}$ & $~0.0^*~$& $~0.0^*~$  \\ 
 \hline
\multicolumn{2}{|c|}{$\alpha _{s}(Q_{0}^2)$}&{$0.365\pm 0.011$}&{$0.362\pm 0.016$} \\
\hline \multicolumn{2}{|c|}{$\chi ^{2}/ndf$}&{$405.38/508=0.798$}&{$559.6/508=1.101$} \\
\hline\hline
\end{tabular}
\end{table}
%
%
\begin{table*}[!htbp]
\caption{\label{tab:g2twist3} Parameter values for the coefficients of the twist-3 corrections at $Q^2=4$ GeV$^2$ obtained in the full scenario.}
\begin{ruledtabular}
\begin{tabular}{lccccc}
      & \textbf{A}  & \textbf{B}  & \textbf{C}  &\textbf{D} &\textbf{E}  \\   \hline\hline
$g_{2,p}^{tw-3}$  & $0.034$ &$0.554 $ & $-0.387 $ &  $-1.17$ &  $0.969$ \\
$g_{2,n}^{tw-3}$  & $0.067$ &$0.106 $ & $-0.448$ &  $0.569$ &  $-0.098$ \\
$g_{2,d}^{tw-3}$  & $0.307$ &$0.117 $ & $-0.210 $ &  $0.657$ &  $-0.083$ \\
\end{tabular}
\end{ruledtabular}
\end{table*}
%
\begin{table*}[!htbp]
\caption{\label{tab:data} Published data points above $Q^2 = 1.0$ GeV$^2$.
Each experiment is given the $x$ and $Q^2$ ranges, the number
of data points for each given target, and the fitted normalization
shifts ${\cal{N}}_i$ (see the text).}
\begin{ruledtabular}
\begin{tabular}{lccccc}
\textbf{Experiment} & \textbf{Ref.} & \textbf{$x$ range}  & \textbf{Q$^{2}$ range {(}GeV$^{2}${)}}  & \textbf{\# of data points} &   \textbf{${\cal N}_{n}$}       \tabularnewline
\hline\hline
E143(p)   &\citep{E143pd}   & 0.031-0.749   & 1.27-9.52 & 28 & 0.9999\\
HERMES(p) & \citep{HERM98}  & 0.028-0.66    & 1.01-7.36 & 39 & 1.0011\\
SMC(p)    &\citep{SMCpd}    & 0.005-0.480   & 1.30-58.0 & 12 & 0.9998\\
EMC(p)    &\citep{EMCp}     & 0.015-0.466   & 3.50-29.5 & 10 & 1.0050\\
E155      &\citep{E155p}    & 0.015-0.750   & 1.22-34.72 & 24 & 1.0189\\
HERMES06(p) &\citep{HERMpd} & 0.026-0.731   & 1.12-14.29 & 51 &0.9990 \\
COMPASS10(p) &\citep{COMP1} & 0.005-0.568   & 1.10-62.10 & 15 &0.9904\\
\multicolumn{1}{c}{$\boldsymbol{g_1^p}$}  &  &  &  &  \textbf{179}  & \\ \hline
E143(d)  &\citep{E143pd}    & 0.031-0.749   & 1.27-9.52    & 28 & 0.9998\\
E155(d)  &\citep{E155d}     & 0.015-0.750   & 1.22-34.79   & 24 & 1.0001\\
SMC(d)   &\citep{SMCpd}     & 0.005-0.479   & 1.30-54.80   & 12 & 1.0000\\
HERMES06(d) & \citep{HERMpd}& 0.026-0.731   & 1.12-14.29   & 51 & 0.9992  \\
Compass05(d)& \citep{COMP2005}& 0.0051-0.4740 &1.18-47.5   & 11 & 0.9980 \\
Compass06(d)& \citep{COMP2006}&0.0046-0.566 & 1.10-55.3    & 15 & 1.0000 \\
\multicolumn{1}{c}{ $\boldsymbol{g_1^d}$} &  &  & & \textbf{141} &    \\ \hline
E142(n)   &\citep{E142n}    & 0.035-0.466   & 1.10-5.50    & 8 & 0.9990 \\
HERMES(n) &\citep{HERM98}   & 0.033-0.464   & 1.22-5.25    & 9 &1.0000\\
E154(n)   &\citep{E154n}    & 0.017-0.564   & 1.20-15.00   & 17 &0.9995 \\
HERMES06(n) &\citep{HERMn}  &  0.026-0.731  & 1.12-14.29   & 51 & 1.0000 \\
Jlab03(n)&\citep{JLABn2003} & 0.14-0.22     & 1.09-1.46    & 4 & 1.0001\\
Jlab04(n)&\citep{JLABn2004} &0.33-0.60      & 2.71-4.8     & 3 &1.0996 \\
Jlab05(n)&\citep{JLABn2005} & 0.19-0.20     &1.13-1.34     & 2 & 1.0353 \\
\multicolumn{1}{c}{$\boldsymbol{g_1^n}$}  &  &  & & \textbf{94} &   \\ \hline
E143(p    &\citep{E143pd}   & 0.038-0.595   & 1.49-8.85    & 12& 0.9999 \\
E155(p   &\citep{E155pdg2}  &0.038-0.780    & 1.1-8.4      & 8 &0.9961\\
Hermes12(p)&\citep{hermes2012g2} &0.039-0.678&1.09-10.35   & 20&0.9992 \\
SMC(p)      &\citep{SMCpg2} & 0.010-0.378    & 1.36-17.07   & 6 &1.0000\\
\multicolumn{1}{c}{$\boldsymbol{g_2^p}$}  &  &  & & \textbf{46} &    \\ \hline
E143(d)     &\citep{E143pd} & 0.038-0.595   & 1.49-8.86    & 12 &1.0001\\
E155(d)    &\citep{E155pdg2}& 0.038-0.780    & 1.1-8.2      & 8 &1.0005\\
\multicolumn{1}{c}{$\boldsymbol{g_2^d}$}  &  &  & & \textbf{20} &   \\ \hline
E143(n)    &\citep{E143pd}  & 0.038-0.595   & 1.49-8.86    & 12& 1.0000\\
E155(n)    &\citep{E155pdg2}&0.038-0.780    &1.1-8.8       & 8 &0.9995\\
E142(n)    &\citep{E142n}   &0.036-0.466    &1.1-5.5       & 8 &1.0000\\
Jlab03(n)  &\citep{JLABn2003}&0.14-0.22     & 1.09-1.46    & 4 &0.9928\\
Jlab04(n)  &\citep{JLABn2004}&0.33-0.60     & 2.71-4.83    & 3 &0.9477\\
Jlab05(n)  &\citep{JLABn2005}&0.19-0.20     & 1.13-1.34    & 2 &0.9888\\
\multicolumn{1}{c}{$\boldsymbol{g_2^n}$}  &  &  & &\textbf{37} &   \\ \hline \hline
\multicolumn{1}{c}{\textbf{Total}}&\multicolumn{5}{c}{~~~~~~~~~~~~~~~~~~~~~~~~~~~~~~~~~~~~~~~~~~~~~~~~~~~~~~~~~~~~~~~~\textbf{517}}
\\
\end{tabular}
\end{ruledtabular}
\end{table*}
\subsection{Overview of data sets}
We use a wide range of polarized deep-inelastic scattering lepton-nucleon data on spin structure functions $g_1$ \cite{E143pd,HERM98,SMCpd,EMCp,E155p,HERMpd,COMP1,E155d,COMP2005,COMP2006,E142n,E154n,HERMn,JLABn2003,JLABn2004,JLABn2005} and $g_2$ \cite{E143pd,E155pdg2,hermes2012g2,SMCpg2,E142n,JLABn2003,JLABn2004,JLABn2005}, which are extracted based on the different nucleon targets of protons, neutrons, and deuterons to extract all PPDFs.

The major properties of these data sets are summarized in Table~\ref{tab:data}, which contains the name of the experimental group, the covered kinematic ranges in $x$ and $Q^2$, the number of available data points, and the fitted normalization shifts ${\cal{N}}_i$.
Our analysis is limited to the region of  $Q^2 \geq 1$ GeV$^2$, to ensure that perturbative QCD is applicable, and $W^2 \ge 3 $ GeV$^2$. The cut on $W^2$ is slightly smaller than in some previous PPDF analyses.

Although most of the $g_2$ data have large errors, we considered them in the fitting procedure. Thus, our results focus on the quality or characteristic of the twist-3 part rather than on their quantity.

\subsection{Method of minimization and error calculation}
$\chi^2(p)$ quantifies the goodness of fit to the data for a set of independent parameters $p$ that specifies the PDFs at $Q_0^2$ \cite{Stump:2001gu}:
\begin{eqnarray} \label{chi2}
\chi_{\mathrm{\rm global}}^{2}(p) &=& \sum_{i=1}^{n^{exp}} \left[\left(\frac{{\cal N}_{i} - 1}{\Delta{\cal N}_{i}}\right)^{2} \right. \nonumber \\ \left.
\right. &+& \left. \sum_{j=1}^{n^{data}}  \left( \frac{{\cal N}_{i} g_{j}^{\rm data} - g^{\rm theory}_{j} (p)} {{\cal N}_{i} ~ \Delta g_{j}^{\rm data}} \right)^{2}\right]\,.
\end{eqnarray}
$n^{exp}$ and $n^{data}$ are the number of individual experimental data sets and corresponding number of data points included in each data set, respectively.
For the $ith$  experiment, each data value $g_{j}^{data}$ with measurement uncertainty $\Delta g_{j}^{data}$ is compared to the corresponding theoretical value $g_{j}^{theor}$. The correlated normalization uncertainty $\Delta {\cal N}_{i}$ is reported for most experiments.
$\Delta {\cal N}_{i}$ is the experimental normalization uncertainty, and ${\cal N}_{i}$ is an overall normalization factor for the data of experiment $i$. We allow for a relative normalization shift ${\cal N}_{i}$ between different data sets within uncertainties $\Delta {\cal N}_{i}$ quoted by the experiments.
The minimization of the above $\chi^2$ function is done using the program MINUIT \cite{James:1994vla}.

\section{Discussion of fit results}
\label{Discussion of fit results}
In this section, we describe our fit, which was performed including target mass corrections to the leading twist contributions and considering higher twist terms. We extract the pure twist-2 and twist-3 contributions along with strong coupling constant.

The standard scenario to extract PDFs from observable is to consider a certain functional form in the leading twist in the $\overline{{\rm MS}}$ scheme as our reference distribution. Their scale dependence is given by the well-known DGLAP evolution equations.
Here, we have performed all $Q^2$ evolutions in Mellin space using the QCD-PEGASUS program \cite{Vogt:2004ns} in the fixed flavor number scheme. The number of active flavors in the splitting functions and Wilson coefficients is fixed at $N_f=3$.

\subsection{Polarized PDFs}
In our QCD analysis, we perform two fitting scenarios to distinguish the effect of target mass corrections and higher twist contribution. These contributions are both considered in the ``full scenario'', while the ``pQCD scenario'' is based on the twist-2 NLO pQCD $g_1$ and twist-2 WW $g_2$ (see  Table~\ref{tab:result}). In the following sections, we indicate full scenario by ``model''.

The $\chi ^{2}$ fit value of the full scenario is smaller than the pQCD scenario, indicating the importance of low-$Q^2$ corrections.
It supports our theoretical framework in which the leading twist part is enriched by TMCs and HT terms.
As shown in Table~\ref{tab:result}, the precision of the extracted PPDFs is essentially enhanced, which is a consequence of above discussed corrections. The strong coupling constant receives corrections of $0.003$ at a scale of $Q_0^2$.

\begin{figure}[!htbp]
\includegraphics[clip,width=0.48\textwidth]{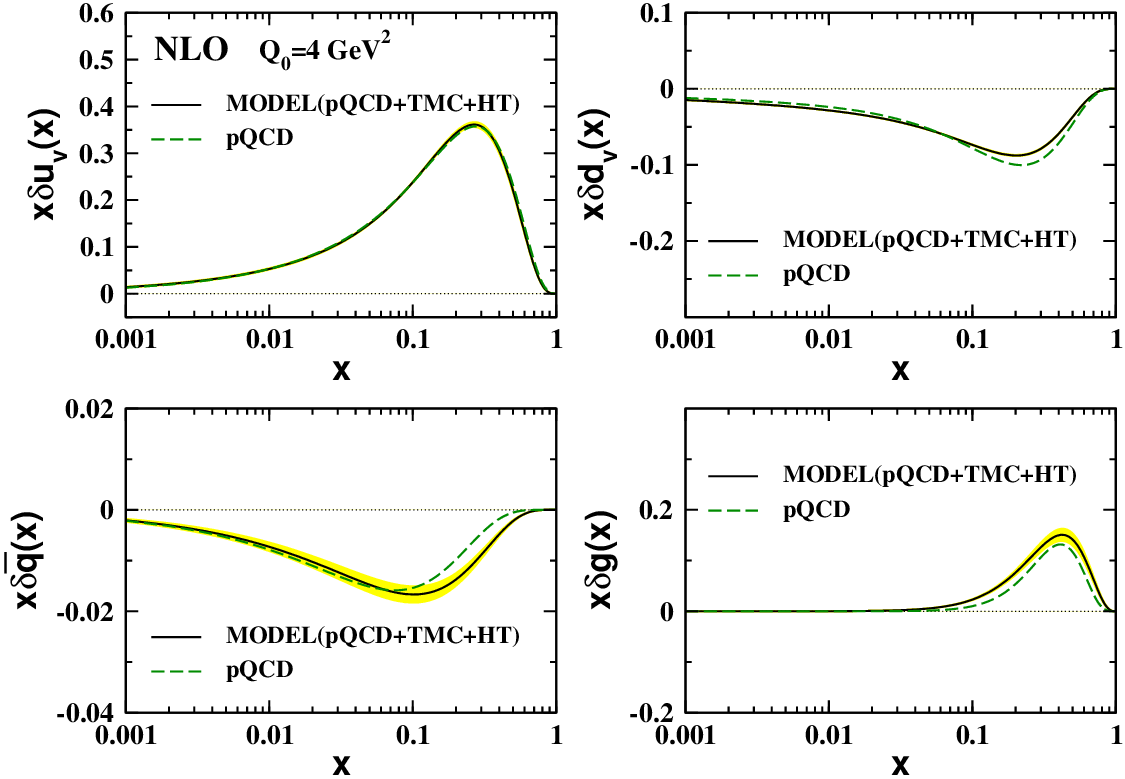}
\caption{{\small The polarized parton distribution at $Q_{0}^{2}=$ 4 GeV$^{2}$ as a function of $x$. Our model is represented by the solid curve, and the pQCD scenario is represented by the dashed curve.  \label{fig:ourPPDF}}}
\end{figure}

\begin{figure}[!htbp]
\includegraphics[clip,width=0.48\textwidth]{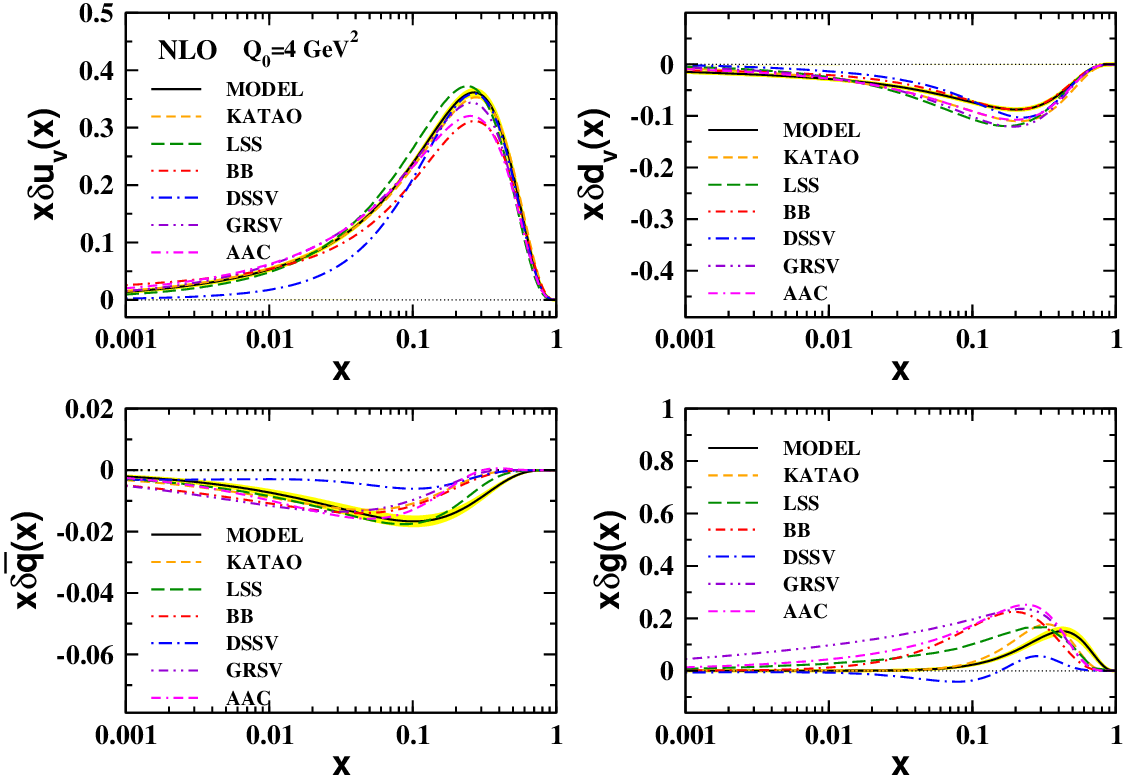}
\caption{{\small  The polarized parton distribution at $Q_{0}^{2}=$ 4 GeV$^{2}$
as a function of $x$. Our fit is the solid curve. Also shown are
the results of KATAO (dashed) \citep{Khorramian:2010qa}, LSS (long dashed) \citep{Leader:2006xc}, BB (dashed dotted) \citep{Blumlein:2010rn}, DSSV (long dashed dotted)
\citep{deFlorian:2008mr}, GRSV (dashed-dotted dotted) \citep{Gluck:2000dy},
and AAC (long dashed-dashed dotted) \citep{Hirai:2008aj}. \label{fig:PPDF}}}
\end{figure}
We compare the PPDFs extracted based on these two scenarios in Fig. \ref{fig:ourPPDF}.
Large-$x$ sea distribution is the most affected part, while $x\delta u_V$ is the least.
In Fig.~\ref{fig:PPDF}, we compared our model with various parameterizations from the literature
\cite{Khorramian:2010qa,Leader:2006xc,Blumlein:2010rn,deFlorian:2008mr,Gluck:2000dy,Hirai:2008aj}.
Most of the fits are comparable. The differences originate from the choice of data sets, the form of PPDF parametrization, and several details of the QCD analysis. For example, the LSS analysis \cite{Leader:2006xc} considered the impact of higher twist corrections on their PPDFs, or the DSSV study \cite{deFlorian:2008mr} included semi-inclusive data in its fitting pass and had different curves for $\bar{u}$, $\bar{d}$, and $\bar{s}$.
\subsection{Polarized structure functions}
\begin{figure*}[!htbp]
\includegraphics[clip,width=0.9\textwidth]{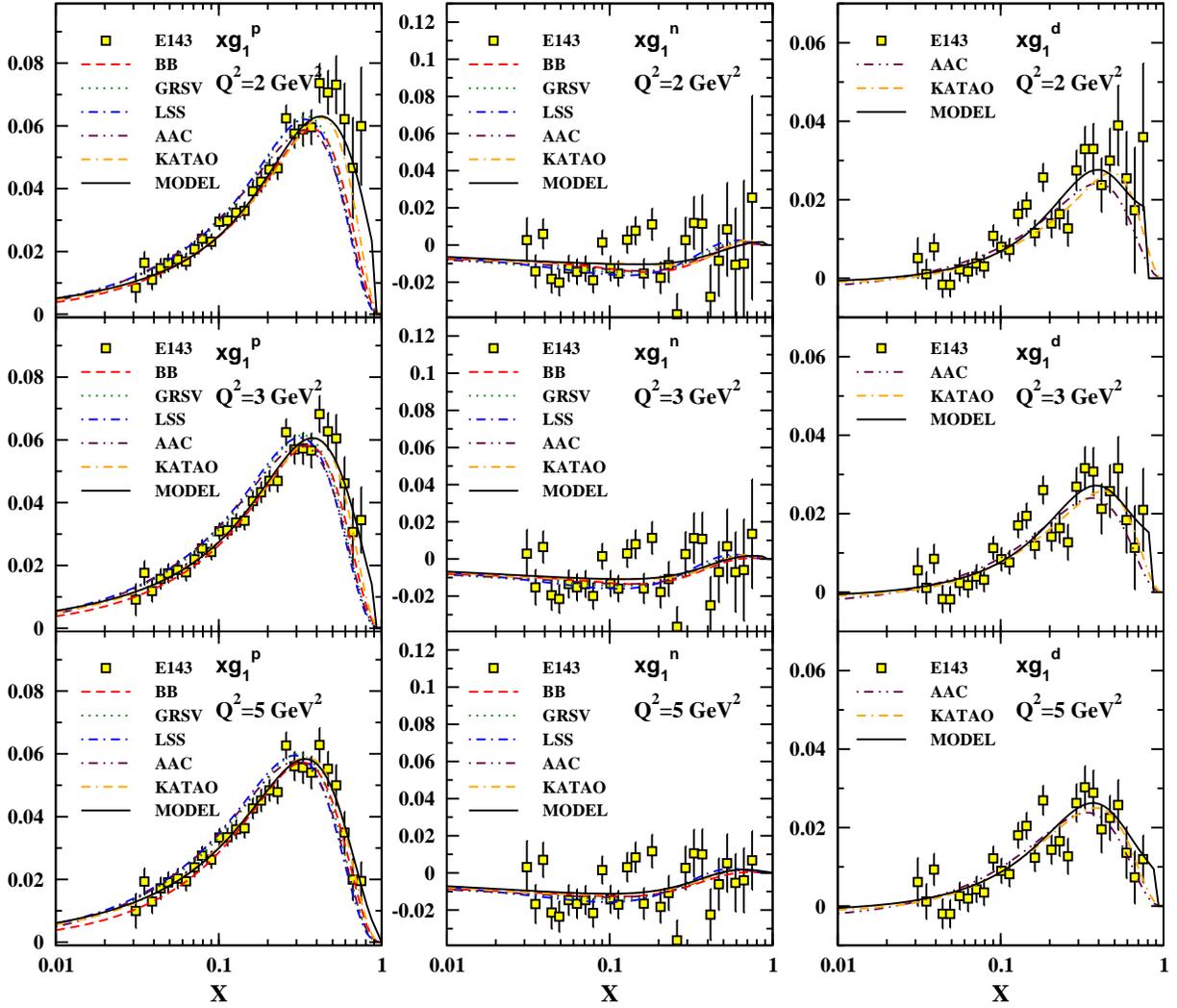}
\caption{{\small The polarized structure functions as a function of $x$ and
for different low values of $Q^{2}$.
Our result (solid curve) is compared with the curves obtained by BB (dashed) \citep{Bluemlein:2002be},
GRSV (dotted) \citep{Gluck:2000dy}, LSS (dashed dotted) \citep{Leader:2005ci},
AAC (dashed-dotted dotted) \citep{Goto:1999by}, and KATAO (dashed-dashed dotted)
\citep{Khorramian:2010qa}. \label{fig:xg1pnd}}}
\end{figure*}

\begin{figure}[!htbp]
\includegraphics[clip,width=0.48\textwidth]{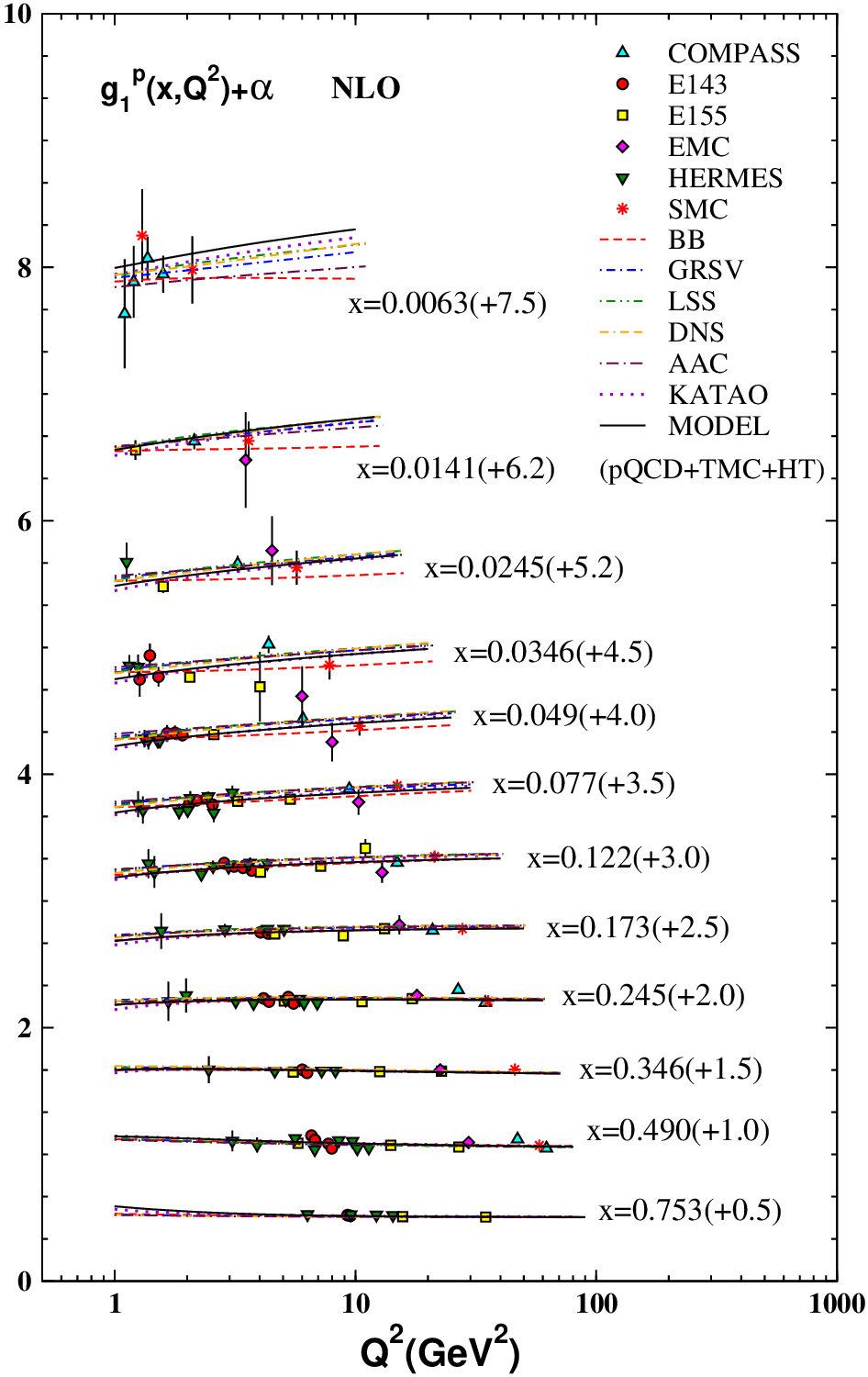}
\caption{{\small The structure function $g_1^p(x,Q^2)$ as a function of $Q^2$ in different intervals of $x$ compared to experimental data. Also shown are the results of BB (dashed) \citep{Bluemlein:2002be},
GRSV (dashed dotted) \citep{Gluck:2000dy}, LSS (dashed-dotted dotted)
\citep{Leader:2005ci}, DNS (dashed-dashed dotted) \citep{deFlorian:2005mw}, AAC (long dashed dotted) \citep{Goto:1999by}, and KATAO (dotted) \citep{Khorramian:2010qa}. To improve legibility, the values of $g_1(x,Q^2)$ have been shifted by the amount of $\alpha$.\label{fig:QCDfit}}}
\end{figure}

\begin{figure}[!htbp]
\includegraphics[clip,width=0.48\textwidth]{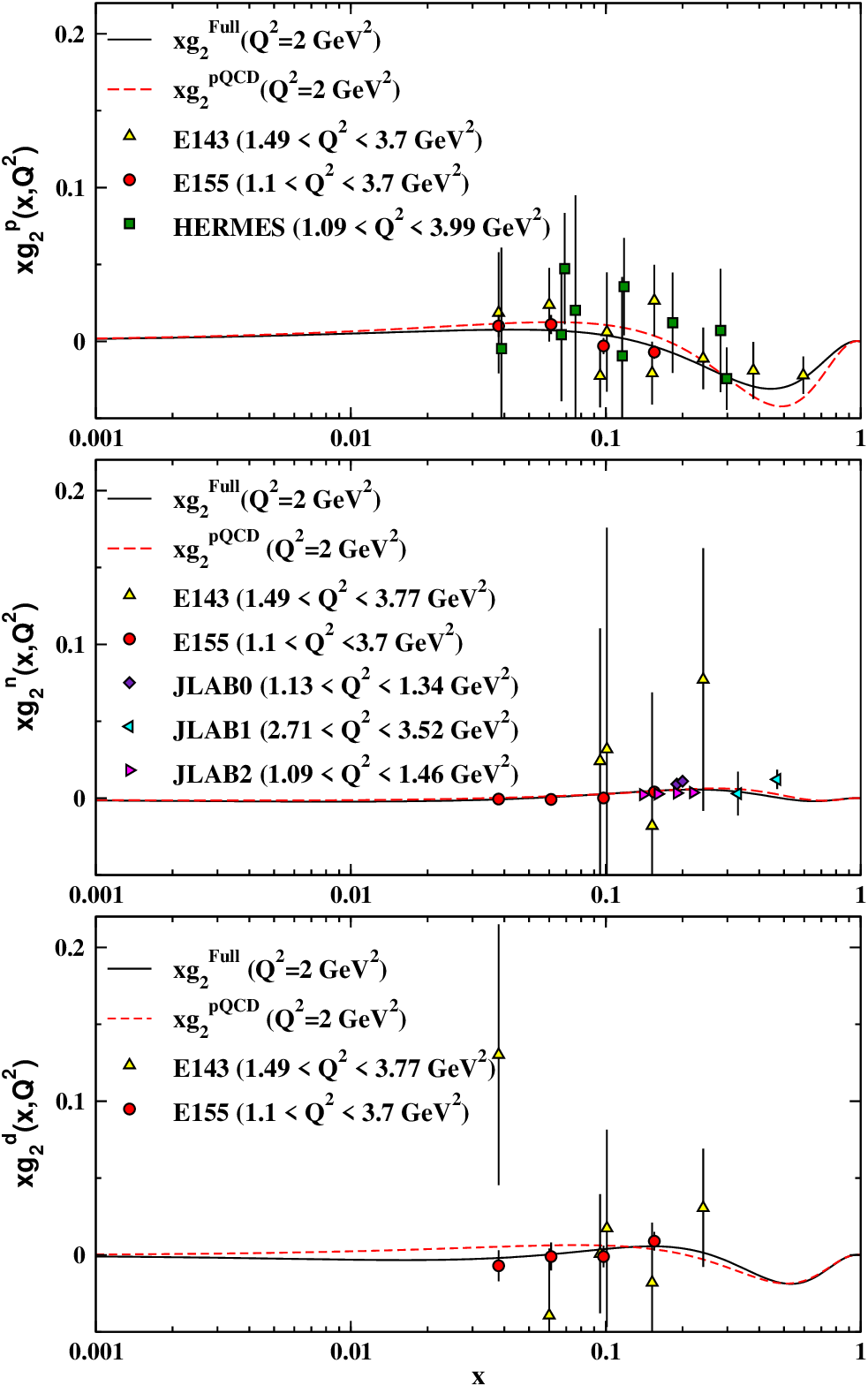}
\caption{{\small The structure function $xg_2$ at $Q^{2}=$ 2 GeV$^{2}$ as a function of $x$ compared to experimental data. \label{fig:xg2versusx}}}
\end{figure}

\begin{figure}[!htbp]
\includegraphics[clip,width=0.48\textwidth]{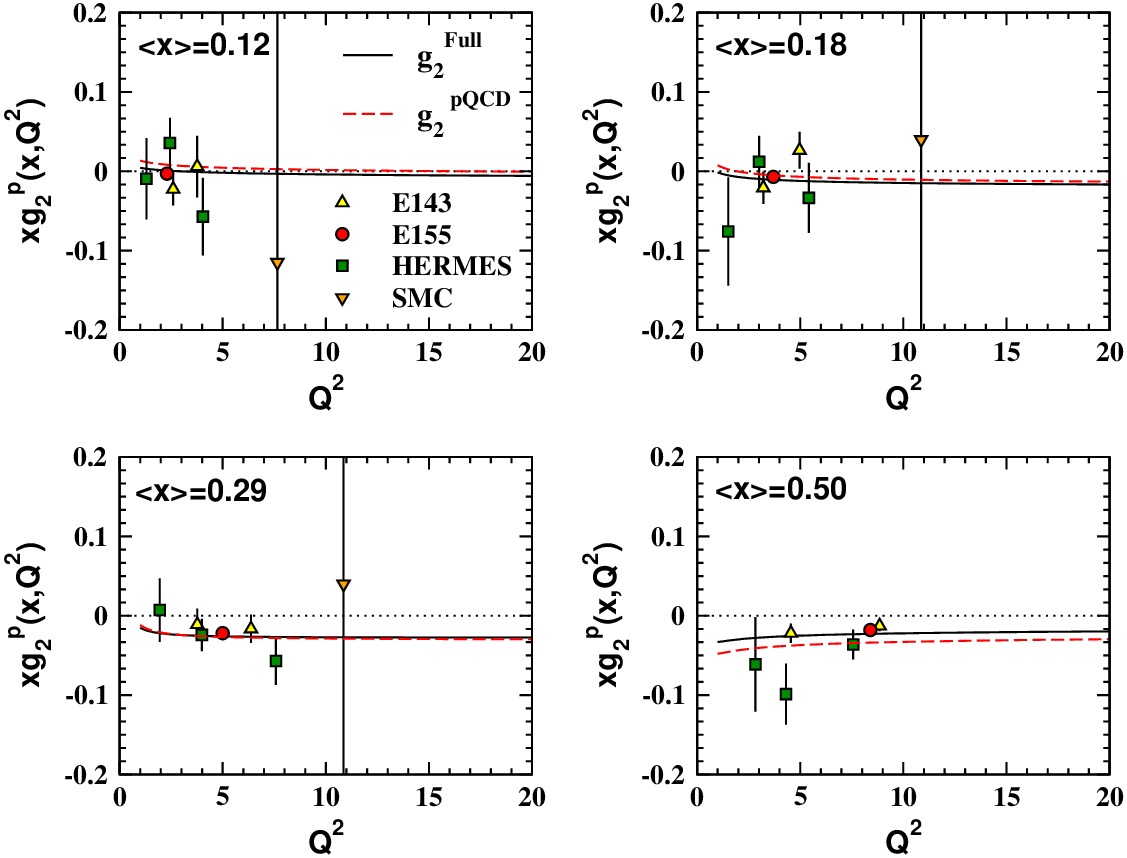}
\caption{{\small The polarized structure function $xg_2^{p}$ as a function of $Q^{2}$ and
for different values of $x$ compared with experimental data. \label{fig:xg2versusQ}}}
\end{figure}

In Fig.~\ref{fig:xg1pnd}, we plot our results for $xg_1^{p,n,d}(x,Q^2)$ as a function of $x$ for low values of $Q^2$. The effect of the $\theta$ function is visible in large $x$. Our curves are well described by the data.
Figure ~\ref{fig:QCDfit} shows our prediction for $g_1^p(x,Q^2)$ as a function of $Q^2$ and for different values of $x$, in comparison with the others \cite{Bluemlein:2002be,Gluck:2000dy,Leader:2005ci,deFlorian:2005mw,Goto:1999by,Khorramian:2010qa}.
The data are well described within errors.
Despite the limited range in $Q^2$, scaling violations of $g_1$ are obviously visible.
The $xg_2$ polarized structure functions for the proton, neutron, and deuteron are shown as a function of $x$ in Fig.~\ref{fig:xg2versusx}. We compare our results based on two scenarios with the experimental data from Refs. \citep{E143pd,E155pdg2,hermes2012g2,SMCpg2,JLABn2003,JLABn2004,JLABn2005}.
In Fig.~\ref{fig:xg2versusQ}, $xg_2^{p}$ as a function of $Q^{2}$ is compared with experimental data \citep{E143pd,E155pdg2,hermes2012g2,SMCpg2}.
\subsection{HT contributions}
\begin{figure}[!htbp]
\includegraphics[clip,width=0.36\textwidth]{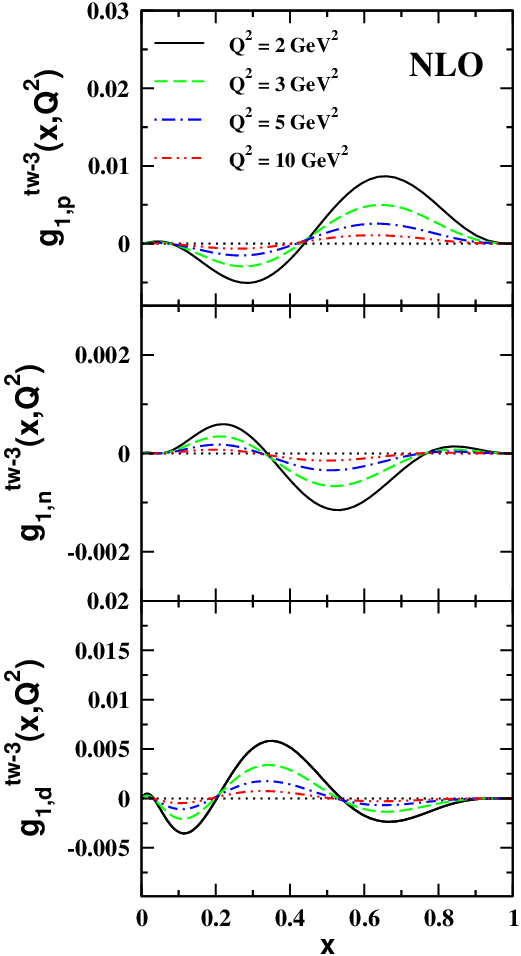}
\caption{{\small The twist-3 contribution of $g_1$ for the proton, neutron, and deuteron as a function of $x$ and
for different values of $Q^{2}$. \label{fig:g1hp-n-d}}}
\end{figure}

\begin{figure}[!htbp]
\includegraphics[clip,width=0.36\textwidth]{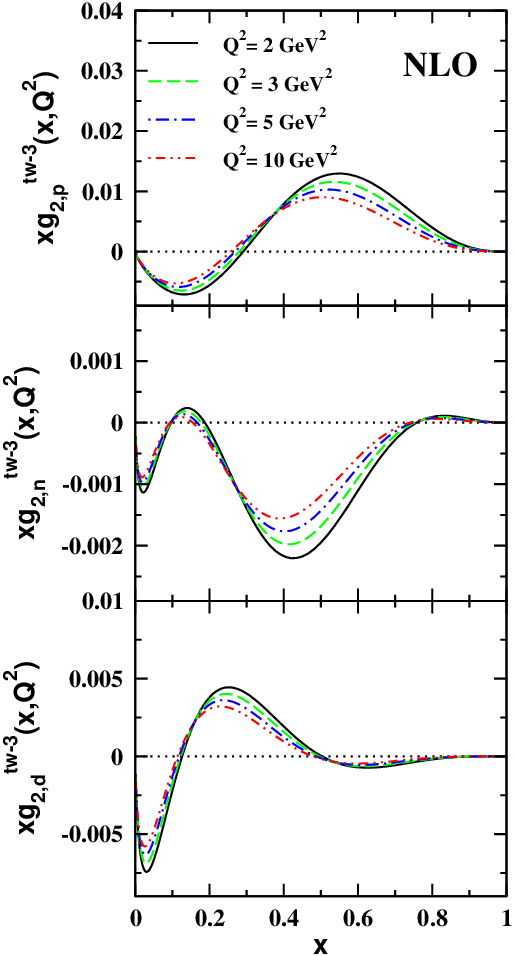}
\caption{{\small The twist-3 contribution of $xg_2$ for the proton, neutron and deuteron as a function of $x$ and
for different values of $Q^{2}$. \label{fig:g2hpnd}}}
\end{figure}

Our results for the twist-3 contribution of both polarized structure functions are shown at $Q^2=1,3,5,10$ in Figs.~\ref{fig:g1hp-n-d} and \ref{fig:g2hpnd}. Note that they vanish with the evolution in the high-$Q^2$ regime.
A significant positive twist-3 modification observes for $g_{2,p}^{tw-3}$ at $x\ge0.3$, which is even larger than the $g_{1,p}^{tw-3}$ modification and as shown in Fig.~\ref{fig:xg2versusx}, cancels some of the negative leading twist contribution.
On the contrary, the $g_{2,n}^{tw-3}$ is approximately zero. A similar result is reported in Ref. \cite{Jimenez-Delgado:2013boa}.

\begin{figure}[!htbp]
\includegraphics[clip,width=0.39\textwidth]{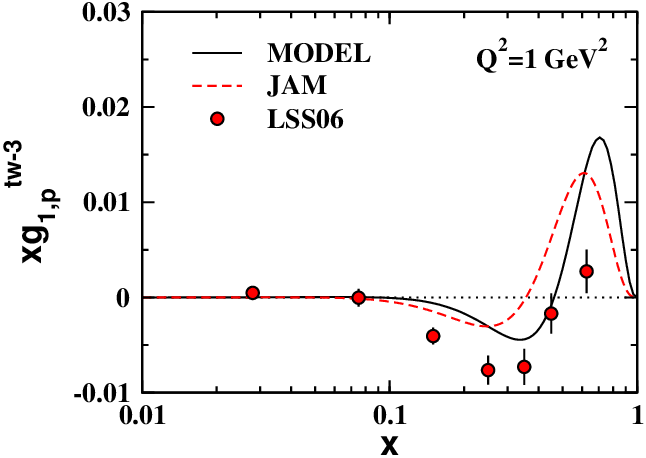}
\caption{{\small The twist-3 contribution to $xg_1$ at $Q^2=1$ GeV$^2$ as a function of $x$ compared to the results of
LSS \cite{Leader:2006xc} and JAM \cite{Jimenez-Delgado:2013boa}. \label{fig:xg1pHT}}}
\end{figure}

\begin{figure}[!htbp]
\includegraphics[clip,width=0.39\textwidth]{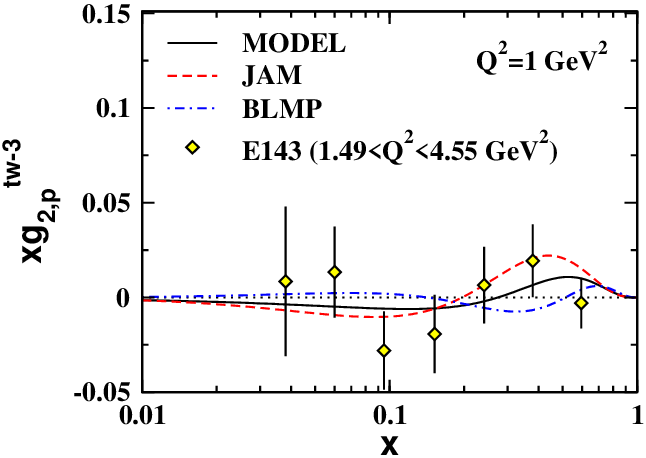}
\caption{{\small The twist-3 contribution to $xg_2$ at $Q^2=1$ GeV$^2$ as a function of $x$.
The current fit is the solid curve. Also shown are the curve based on JAM \cite{Jimenez-Delgado:2013boa} (dashed), BLMP \cite{Braun:2011aw} (dashed dotted) and E143 experimental data. \label{fig:xg2pHT}}}
\end{figure}

In Fig. \ref{fig:xg1pHT}, we compare our result on twist-3 contributions to $g_1^p$ with those obtained by LSS \cite{Leader:2006xc} and JAM \cite{Jimenez-Delgado:2013boa}. The LSS group extracted effective HT in a model-independent way from experimental data corresponding to seven $x$ bins. However, the logarithmic $Q^2$ dependence of the twist-3 parts is neglected.
The JAM global NLO analysis is based on a direct fit of the measured longitudinal and transverse asymmetries.
In Fig. \ref{fig:xg2pHT}, the twist-3 contribution to $xg_2$ is compared with the JAM \cite{Jimenez-Delgado:2013boa} and BLMP models \cite{Braun:2011aw} along with the E143 experimental data \citep{E143pd}. Our results are comparable with  theoretical and phenomenological predictions.

\subsection{Strong coupling constant}
In our analysis, the strong coupling constant is considered as a free parameter. 
 Although some phenomenological groups, such as LSS \cite{Leader:2006xc,Leader:2010rb} or AKS \cite{Arbabifar:2013tma}, fix $\alpha_s$ close to the updated Particle Data Group average, other phenomenological groups, such as BB \cite{Bluemlein:2002be, Blumlein:2010rn} or KATAO \cite{Khorramian:2010qa}, extract it as a free parameter.
 To take into account its correlation with other parameters, the strong coupling constant extracted simultaneously with the PPDFs and higher twist terms.
 We achieve the value of $\alpha_s(Q_0^2)=0.365\pm 0.011$ in our model. This value is closely related to the gluon distribution, which drives the QCD evolution.
 
The scale dependence of the running coupling constant at NLO is precisely given in terms of $a_s(Q_0^2)$ by
\begin{eqnarray}
\frac{1}{a_s(Q^2)}&=&\frac{1}{a_s(Q_0^2)}+\beta_0\ln\left(\frac{Q^2}{Q_0^2}\right)\nonumber\\
&&- b_1
\ln\left\{\frac{a_s(Q^2)[1+b_1a_s(Q_0^2)]}{a_s(Q_0^2)[1+b_1a_s(Q^2)]}\right\}~.
\label{as}
\end{eqnarray}
Here $a_s=\frac{\alpha_s}{4\pi}$ and $b_1=\frac{\beta_1}{\beta_0}$. The $\beta$ functions are known up to N$^3$LO and depend on the number of active flavors \cite{Botje:1999dj,Furmanski:1981cw,Larin:1993tp,vanRitbergen:1997va}.
Rescaling the coupling constant to the Z boson mass, we obtain $\alpha_s(M_Z^2)=0.1136\pm 0.0012$, which is comparable with the current world average $\alpha_s(M_Z^2)=0.1184\pm 0.0007$.
\section{Polarized Structure function of $^{3}\rm He$ and $^{3}\rm H$}
\label{Polarized Structure function of $^{3}He$ and $^{3}H$}
${^{3}\rm He}$ and ${^{3}\rm H}$ are two of simplest nuclei, which consist of 2(1) protons and 1(2) neutron. Because of different nuclear effects, protons and neutrons inside the nuclei are different from those in free space. The most important effects are spin depolarization, nuclear binding, and Fermi motion, which are available in the framework of the convolution approach \citep{Bissey:2001cw}. In this approximation, $g_{1}^{^{3}\rm He}$ and $g_{1}^{^{3}\rm H}$ can be interpreted as the convolution of $g_{1}^{p}$ and $g_{1}^{n}$ with the spin-dependent nucleon light-cone momentum distributions $\Delta f_{(^{3}\rm He,^{3}\rm H)}^{N}(y)$ as follows \cite{Schulze:1992mb,Schulze:1997rz}:
\begin{eqnarray}
g_{1}^{^{3}\rm He}(x,Q^{2}) & = & \int_{x}^{3}\frac{dy}{y}\Delta f_{^{3}\rm He}^{n}(y)g_{1}^{n}(\frac{x}{y},Q^{2})\nonumber \\
 & + & 2\int_{x}^{3}\frac{dy}{y}\Delta f_{^{3}\rm He}^{p}(y)g_{1}^{p}(\frac{x}{y},Q^{2})\nonumber \\
 & - & 0.014[g_{1}^{p}(x,Q^{2})-4g_{1}^{n}(x,Q^{2})]\;,\label{eq:g1He}\end{eqnarray}
 \begin{eqnarray}
g_{1}^{^{3}\rm H}(x,Q^{2}) & = & 2\int_{x}^{3}\frac{dy}{y}\Delta f_{^{3}\rm H}^{n}(y)g_{1}^{n}(\frac{x}{y},Q^{2})\nonumber \\
 & + & \int_{x}^{3}\frac{dy}{y}\Delta f_{^{3}\rm H}^{p}(y)g_{1}^{p}(\frac{x}{y},Q^{2})\nonumber \\
 & + & 0.014[g_{1}^{p}(x,Q^{2})-4g_{1}^{n}(x,Q^{2})]\;.\label{eq:g1H}\end{eqnarray}
 $\Delta f_{(^{3}\rm He,^{3}\rm H)}^{N}(y)$ is the probability to find $N$ in the $(^{3}\rm He,^{3}\rm H)$ with a given fraction of the total momentum $y$.
The light-cone momentum distributions for the proton and neutron in the three-nucleon system is determined. Concerning isospin symmetry, $f_{^{3}\rm He}^{p} (f_{^{3}\rm He}^{n})$ and $f_{^{3}\rm H}^{n} (f_{^{3}\rm H}^{p})$ are equal.

\begin{figure}[!htbp]
\includegraphics[clip,width=0.39\textwidth]{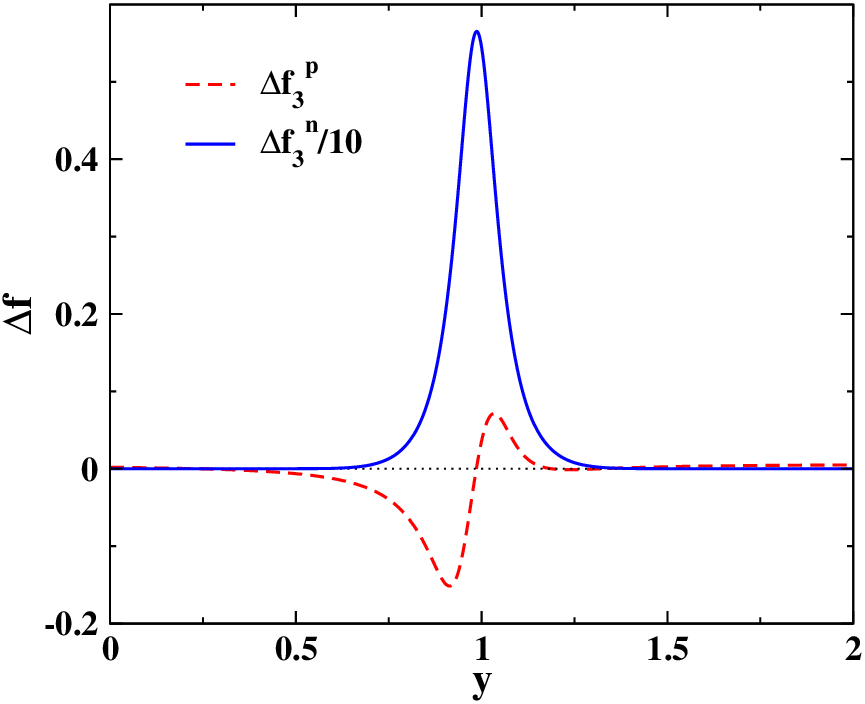}
\caption{{\small The polarized light-cone distribution functions for proton and neutron in $^3$He.
\label{fig:lightcone}}}
\end{figure}

 We used the results of  Refs.~\citep{Bissey:2001cw,Bissey:2000ed,Afnan:2003vh} as
\begin{equation}
\Delta f_{^{3}\rm He}^{n}(y)=\frac{a^{n}e^{-\frac{0.5(1-d^{n})(-b^{n}+y)^{2}}{{(c^{n})}^{2}}}}{1+\frac{d^{n}(-b^{n}+y)^{2}}{{(c^{n})}^{2}}}~,\label{eq:f3HeN}\end{equation}
 \begin{equation}
\Delta f_{^{3}\rm He}^{p}(y)=\frac{\sum_{i=0}^{4}a_{i}^{p}U_{i}(y)}{\sum_{i=0}^{4}b_{i}^{p}U_{i}(y)}~.\label{eq:f3HeP}\end{equation}
Here, $U_n(y)$ is the Chebyshev polynomial of the second kind.
The numerical multipliers of the above equations are discussed in Ref. \citep{Khorramian:2010qa}.
 Figure ~\ref{fig:lightcone} represents our polarized light-cone distribution which is written based on numerical results of Ref.~\citep{Afnan:2003vh}.

\begin{figure}[!htbp]
\includegraphics[clip,width=0.38\textwidth]{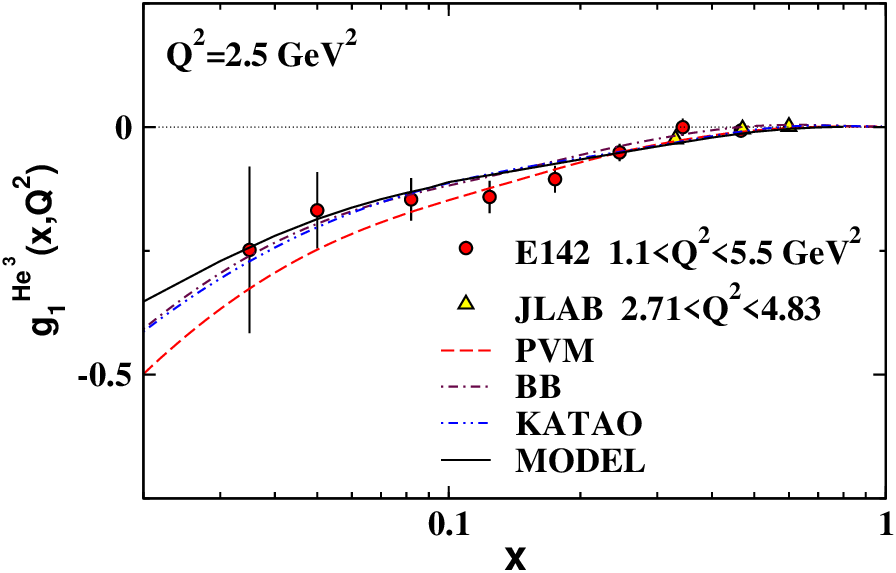}
\caption{{\small $g_1^{He^3}$ for fixed $Q^{2}=2.5$ GeV$^{2}$.
The current fit is the solid curve. Also shown are the curves based on the polarized valon model (PVM) (dashed) \citep{Atashbar Tehrani:2007be}, BB (dashed dotted) \citep{Bluemlein:2002be} and KATAO (dashed-dotted dotted) \cite{Khorramian:2010qa}. \label{fig:g1he3}}}
\end{figure}

\begin{figure}[!htbp]
\includegraphics[clip,width=0.38\textwidth]{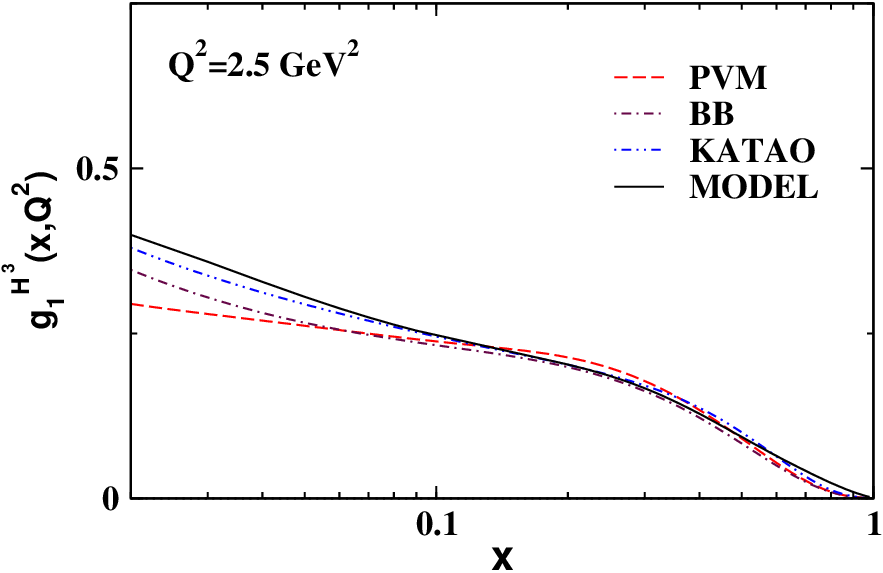}
\caption{{\small $g_1^{H^3}$ for fixed $Q^{2}=2.5$ GeV$^{2}$.
The current fit is the solid curve. Also shown are the curves based on the polarized valon model (PVM) (dashed) \citep{Atashbar Tehrani:2007be}, BB (dashed dotted) \citep{Bluemlein:2002be} and KATAO (dashed-dotted dotted) \cite{Khorramian:2010qa}. \label{fig:g1H3}}}
\end{figure}

Our results for $g_{1}^{^{3}\rm He}$ and $g_{1}^{^{3}\rm H}$ are compared with BB~\citep{Bluemlein:2002be}, PVM~\citep{Atashbar Tehrani:2007be}, and KATAO~\cite{Khorramian:2010qa} in Figs.~\ref{fig:g1he3} and \ref{fig:g1H3}.
\section{Sum rules}
\label{Sum rules}
Parton distribution functions and structure functions follow a series of sum rules.
These sum rules, which are based on the moments of structure functions, provide an opportunity to test QCD.
Moments of structure functions contain valuable information about the total momentum fraction carried by partons or the total contribution of parton helicities to the spin of nucleon in unpolarized or polarized cases. Ellis--Jaffe \cite{Ellis:1973kp} and Bjorken \citep{Bjorken:1966jh} sum rules are based on first moment of $g_1$. The Burkhard--Cottingham \cite{Burkhardt:1970ti} sum rule focuses on the first moment of $g_2$. Moreover, moments of $g_{1,2}$ can be related to matrix elements operators via the OPE.
All these important sum rules are briefly discussed in following.
\subsection{Twist-3 contributions to polarized nucleon structure functions sum rule}
The OPE sum rule relates the moments of $g_1$ and $g_2$ at fixed $Q^2$ to the twist-2 and twist-3 reduced matrix elements of spin-dependent operators in the nucleon, $a_n$ and $d_n$ \cite{Jaffe:1990qh},
\begin{eqnarray}
\Gamma_{1}^{n}= \int_0 ^1 x^n g_1(x,Q^2) dx&=&\frac{a_n}{2},~~~n=0,2,4,...\nonumber \\
\Gamma_{2}^{n}= \int_0 ^1 x^n g_2(x,Q^2) dx&=&\frac{1}{2}\frac{n}{n+1}(d_n-a_n),\nonumber \\
~~~n&=&2,4,...
\end{eqnarray}

For the first moment of $g_2$ ($\int_0 ^1 g_2(x,Q^2) dx$), the OPE does not define any sum rule. But Burkhardt and
Cottingham have derived this value from virtual Compton scattering dispersion relations, which will be discussed in the next part.

The twist-3 matrix elements,
\begin{eqnarray}
d_{n}(Q^2)&=&2 \int_0 ^1 x^n (\frac{n+1}{n})\bar g_2(x,Q^2)dx,\nonumber \\
~~~n&=&2,4,6,...
\end{eqnarray}
measure deviations of $g_2$ from $g_2^{ww}$ term [see Eq. (\ref{WW})].

Having a number of theoretical \cite{Song:1996ea,mit91,bag95,qcdsr1,qcdsr2,lat95} and experimental \cite{Kuhn:2008sy,E143pd} nonzero predictions for $d_2$, which indicate on the role of twist-3 contribution, makes the study of $g_2$ specifically exciting. In Tables \ref{tab:twist2} and \ref{tab:twist3}, we quote theoretical and experimental values for the twist-2 and twist-3 matrix elements for the proton, neutron, and deuteron together with our results.
This remarkably nonzero value for $d_2$ indicates the importance of considering higher twist approximation.
The accuracy of the current data is not sufficient enough to specify model precision.
%
%
\begin{table*}[!htbp]
\caption{\label{tab:twist2} Comparison of theoretical and experimental results for the reduced twist-2 matrix element proton, neutron, and deuteron.}
\begin{ruledtabular}
\begin{tabular}{lccccc}
&Ref.&\textbf{$Q^2$ [GeV$^{2}$]} & \textbf{$a^{p}_2$} & \textbf{$a^{n}_2$} & \textbf{$a^{d}_2$} \tabularnewline
\hline\hline
MODEL      && $5$ & $2.22 \times 10^{-2}$                   & $-3.6\times 10^{-4}$  & $9.97\times 10^{-3}$         \\
CM bag model &\cite{Song:1996ea}&$5$ & $2.10\times 10^{-2}$&$-1.86\times 10^{-3}$  & $8.74\times 10^{-3}$        \\
Lattice QCD &\cite{lat95}&$4$  & $(3.00\pm 0.64)\times 10^{-2}$& $-(2.4\pm 4.0)\times10^{-3}$ & $(13.8\pm 5.2)\times 10^{-3}$  \\
E143 &\cite{E143pd} &$5$  &$(2.48\pm 0.20)\times 10^{-2}$ & $-(4.8\pm 3.2)\times 10^{-3}$& $(9.2\pm 1.6)\times 10^{-3}$\\
\end{tabular}
\end{ruledtabular}
\end{table*}
%
%
\begin{table*}[!htbp]
\caption{\label{tab:twist3} Comparison of theoretical and experimental results for the reduced twist-3 matrix element proton, neutron, and deuteron.}
\begin{ruledtabular}
\begin{tabular}{lccccc}
 & Ref. &\textbf{$Q^2$ [GeV$^{2}$]} &\textbf{$d^{p}_2$}& \textbf{$d^{n}_2$}& \textbf{$d^{d}_2$} \tabularnewline\hline\hline
MODEL      &&$5$& $0.58 \times 10^{-2}$          & $-0.7 \times 10^{-3}$          & $0.6\times 10^{-3}$          \\
JAM model    &\cite{Jimenez-Delgado:2013boa}&$5$  & $1.1\times 10^{-2}$          & $2 \times 10^{-3}$       &  -         \\
CM bag model &\cite{Song:1996ea}&$5$  & $1.74\times 10^{-2}$          & $-2.53 \times 10^{-3}$       & $6.79\times 10^{-3}$         \\
MIT bag model &\cite{mit91,bag95}&$1$ &$1.0\times 10^{-2}$            & $0$                          & $5.0\times 10^{-3}$          \\
QCD sum rule &\cite{qcdsr1}&$1$ & $-(0.6\pm 0.3)\times 10^{-2}$ & $-(30\pm10)\times 10^{-3}$   &$-(17\pm 5)\times 10^{-3}$    \\
QCD sum rule&\cite{qcdsr2}& $1$ & $-(0.3\pm 0.3)\times 10^{-2}$ & $-(25\pm10)\times 10^{-3}$   & $-(13\pm 5)\times 10^{-3}$   \\
Lattice QCD &\cite{lat95}&$4$   & $-(4.8\pm 0.5)\times 10^{-2}$ & $-(3.9\pm2.7)\times 10^{-3}$ & $-(22\pm 6)\times 10^{-3}$   \\
Combined E155 with SLAC data&\cite{Kuhn:2008sy}& $5$ & $(0.32\pm 0.17)\times 10^{-2}$&$(0.79\pm 0.48)\times 10^{-2}$ &     -                         \\
E143 &\cite{E143pd}&$5$                 & $(0.58\pm 0.50)\times 10^{-2}$& $(5.0\pm 21.0)\times 10^{-3}$& $(5.1\pm 9.2)\times 10^{-3}$ \\
\end{tabular}
\end{ruledtabular}
\end{table*}
\subsection{Burkhardt--Cottingham  sum rule}
The first moment of $g_2(x,Q^2)$ follows the Burkhardt--Cottingham (BC) sum rule for all $Q^2$ \cite{Burkhardt:1970ti}:
\begin{equation}
\int_{0}^{1}dx[g_{2}(x,Q^2)]=0~.
\end{equation}
Its validity depends on the lack of singularities for $g_2$ at $x \rightarrow 0$.
Note that this sum rule would automatically be satisfied in twist 2.
Therefore, the presence of higher twist contributions can be concluded from the sum rule violation \cite{hermes2012g2}. In Table~\ref{tab:BC}, our result for this sum rule at $Q^2=5$ GeV$^2$ is compared  with experimental results \cite{E143pd,E155pdg2,hermes2012g2}.
Any conclusion relies on the behavior of $g_2$ at low $x$, which is not accurately known up to now.
%
\begin{table*}[!htbp]
\caption{\label{tab:BC} Comparison of the result of the Burkhardt--Cottingham sum rule with experimental data in $Q^2=5$ GeV$^2$.
 Our result is calculated in two different $x$ ranges.}
\begin{ruledtabular}
\begin{tabular}{lcccc}
      & \textbf{E143} \cite{E143pd}  & \textbf{E155} \cite{E155pdg2}  & \textbf{HERMES2012} \cite{hermes2012g2}  &\textbf{MODEL}   \\
      & $0.03 \le x \le 1 $  & $0.02 \le x \le 0.8 $   & $0.023 \le x \le 0.9$  & $0 (0.02) \le x \le 1 (0.9)$\\
            \hline\hline
$\int g_{2}^{p}(x,Q^2)dx$  & $-0.014 \pm 0.028$ &$-0.044 \pm 0.008 \pm 0.003$ & $0.006 \pm 0.024 \pm 0.017$ &  $-0.008 (-0.012)$ \\
$\int g_{2}^{d}(x,Q^2)dx$  & $-0.034 \pm 0.082$ &$-0.008 \pm 0.012 \pm 0.002$ &-  & $-0.003(-0.006)$ \\
\end{tabular}
\end{ruledtabular}
\end{table*}
%
\subsection{Bjorken sum rule}
The Bjorken sum rule \citep{Bjorken:1966jh} relates the integral over all $x$ at fixed $Q^2$ of the difference between the proton and neutron polarized structure function to the neutron beta decay coupling constant. This sum rule can be explicitly concluded for the $^{3}$He--$^{3}$H system. Considering the ratio of these two relations, one gets \citep{Bissey:2001cw}
\begin{equation}
\frac{\int_{0}^{3}[g_{1}^{^{3}H}(x,Q^{2})-g_{1}^{^{3}He}(x,Q^{2})]dx}{\int_{0}^{1}[g_{1}^{p}(x,Q^{2})-g_{1}^{n}(x,Q^{2})]dx}=\frac{\tilde{g}_{A}}{g_{A}}=0.956\pm0.004~.
\label{eq:ratio}
\end{equation}
We achieved the value of 0.974 for the above ratio.
\subsection{Ellis--Jaffe sum rule}
The first moment of $g_1$ for the proton and neutron was calculated by Ellis--Jaffe sum rules, 
\begin{eqnarray}
\int_{0}^{1}g_{1}^{p}(x,Q^{2})dx&=&\frac{g_A}{12}(1.78)~,\nonumber\\
\int_{0}^{1}g_{1}^{n}(x,Q^{2})dx&=&\frac{g_A}{12}(-0.22)~,
\label{Ellis-Jaffe}
\end{eqnarray}
where $g_A=1.248\pm0.010$ \cite{Ellis:1973kp}.
The $\alpha_s^3$ corrections to these sum rules were calculated in Ref. \cite{Larin:1997qq}.
The Ellis--Jaffe sum rules are not as fundamental as the Bjorken sum rule since they are derived based on a model-dependent assumption that strange quarks do not contribute to the asymmetry.
However, they teach us about the spin structure of the nucleon.
Our result for this sum rule, in the $x$ range of $0.021\leq x \leq 0.9$, is compared with experimental measurements \citep{E143pd,SMCpg2,HERMpd} in Table~\ref{tab:EllisJaffe}.
%
\begin{table*}[!htbp]
\caption{\label{tab:EllisJaffe} Comparison of the result of Ellis--Jaffe sum rule with experimental data in $Q^2=5$ GeV$^2$. Hermes \cite{HERMpd} results are measured in the region $0.021\leq x \leq 0.9$. }
\begin{ruledtabular}
\begin{tabular}{lccccc}
      & \textbf{E143} \cite{E143pd}   & \textbf{SMC} \cite{SMCpg2}  & \textbf{HERMES}  \cite{HERMpd}   & \textbf{MODEL}   \tabularnewline
\hline\hline
$\int g_{1}^{p}(x,Q^2)dx$ & $0.129 \pm 0.003 \pm 0.010$   & $0.132 \pm 0.017 $  & $0.121 \pm 0.009$ & $0.121\pm 0.002$\\
$\int g_{1}^{n}(x,Q^2)dx$ & $-0.034 \pm 0.007 \pm 0.016$  & $-0.048 \pm $ 0.022  &$-0.027 \pm 0.009$  & $-0.027\pm 0.004$\\
$\int g_{1}^{d}(x,Q^2)dx$ & $0.044 \pm 0.003 \pm 0.006$   & $0.039 \pm 0.008 $  & $0.044 \pm 0.003$ & $0.034\pm 0.003$\\
\end{tabular}
\end{ruledtabular}
\end{table*}
\subsection{Efremov--Leader--Teryaev sum rule}
This sum rule involves only the valence contributions of the polarized structure functions $g_{1,2}$ \cite{Efremov:1996hd}:
\begin{equation}
\int_{0}^{1}dxx[g_{1}^{V}(x)+2g_{2}^{V}(x)]=0~.
\end{equation}
Assuming the isospin symmetry of the sea quark distribution, the sum rule takes a form $\int_{0}^{1}dxx[g_{1}^{p}(x)+2g_{2}^{p}(x)-g_{1}^{n}(x)-2g_{2}^{n}(x)]=0$.
It holds under the presence of target mass corrections \cite{Blumlein:1998nv}. We achieved the amount of $1.78 \times 10^{-5}$ at $Q^2=5$ GeV$^2$, which is consistent with zero. The value of $ -0.013 \pm 0.008 \pm 0.002$ is reported by E155 \cite{E155pdg2} at the same $Q^2$.
\subsection{First moment}
The spin contribution of parton $i$ to the nucleon spin can be found by its first moment integral
$\Delta q_{i}(Q^2)=\int_{0}^{1}dx\delta q_{i}(x,Q^2)$.
This is why there are universal efforts to determine the $\delta q_{i}(x,Q^2)$ from different experimental data.
In Table~\ref{tab:firstmoment}, we present the values for the first moments of the polarized quark and gluon extracted from our model at $Q_0^2=4$ GeV$^2$. They are compared with recent fit results of DSSV08 \cite{deFlorian:2008mr} (DIS, SIDIS, and RHIC), BB10 \cite{Blumlein:2010rn} (DIS data), LSS10 \cite{Leader:2010rb} (DIS and SIDIS data), AAC08-Set A \cite{Hirai:2008aj} (DIS), and NFRR12 \cite{Nocera:2012hx} (DIS data). The values of $\Delta \Sigma$ are almost comparable, while different $\Delta g$ are reported.
%
\begin{table*}[!htbp]
\caption{\label{tab:firstmoment}First moments of the polarized singlet-quark $\Delta \Sigma(Q^2)=\sum_{i}\int_{0}^{1}dx[\delta q_{i}(x)+\delta \bar q_{i}(x)]$ and gluon distributions at the scale $Q_0^2=4$ GeV$^2$ in the $\overline{{\rm MS}}$ scheme (only AAC08 \cite{Hirai:2008aj} results are given at $Q_0^2=1$ GeV$^2$).}
\begin{ruledtabular}
\begin{tabular}{ccccccc}
      & \textbf{DSSV08} \cite{deFlorian:2008mr}  & \textbf{BB10} \cite{Blumlein:2010rn} & \textbf{LSS10} \cite{Leader:2010rb}& \textbf{AAC08}\cite{Hirai:2008aj} & \textbf{NFRR12} \cite{Nocera:2012hx}& \textbf{MODEL}     \tabularnewline
\hline\hline
$\Delta \Sigma (Q^2)$  & $0.25 \pm 0.02$ & $0.19 \pm  0.08$ & $0.21 \pm 0.03$ & $0.24 \pm 0.07$ & $0.31 \pm 0.10$ &$0.223 \pm 0.036$\\
$\Delta g (Q^2)$   & $-0.10 \pm 0.16$ & $0.46 \pm 0.43$ & $0.32 \pm 0.19$ & $0.63 \pm 0.81$ & $-0.2 \pm 1.4$ &$0.201 \pm 0.044$ \\
\end{tabular}
\end{ruledtabular}
\end{table*}

\section{Conclusions}
\label{Conclusions}
We have carried out a NLO QCD analysis to the polarized structure functions data $g_1$ and $g_2$.
During the analysis, we considered TMCs and HT effects to extract the PPDFs inside the nucleon.
The strong coupling constant and twist-3 part of the $g_{1}$ and $g_{2}$ are simultaneously determined along with them.
The TMCs are calculated explicitly from the leading twist perturbative polarized structure function within the OPE. 
In contrast to most previous PPDFs studies that neglected the scale dependence of the HT contributions, our model considers the $Q^2$ evolution of the HT terms. This strategy leads to extracting more precise PPDFs with smaller uncertainty bands.
We report smaller $\chi^2$ for our full scenario including low $Q^2$ corrections. This progress demonstrates a clear preference of the data for the existence of their effects. The strong coupling constant also receives a small correction.
Having extracted the polarized PDFs, we estimated the nuclear structure function of $^3\rm He$ and $^3\rm H$.
Finally, we computed the moments of PPDFs and structure functions and discussed about the sum rules.
We found good agreement with the observables, and our outcomes were agreeable with other results from the literature.
More accurate data are required to conclude final determination.

Having analyzed all the polarized inclusive DIS data on $g_1$ and $g_2$, we examined the efficiencies of our method on polarized semi-inclusive reactions (SIDIS) to calculate quark and antiquark densities individually. This work is in progress.

\section{Acknowledgments}
The authors appreciate E.~Leader for reading the manuscript of this paper, fruitful suggestions, and critical remarks.
S.~T. thanks the CERN TH-PH division for its hospitality where a portion of this work was performed.
A.~N.~K. is grateful to the Physics Department of Southern Methodist University for its hospitality.
S.~T. and A.~N.~K. acknowledge the School of Particles and Accelerators, Institute
for Research in Fundamental Sciences (IPM), for financially supporting this project.
\section*{Appendix: FORTRAN code }
A \texttt{FORTRAN} package containing our PPDFs as well as the polarized structure functions $g_{1,2}(x,Q^{2})$ together with the $g_2^{tw-3}$ contribution for the proton, neutron, and deuteron is available in \texttt{http://particles.ipm.ir/links/QCD.htm} or can be obtained via Email from the authors. These functions are interpolated
using cubic splines in $Q^{2}$ and a linear interpolation in $\log\,(Q^{2})$. The
package includes an example program to illustrate the use of the routines.


\end{document}